\documentclass{jpsj3}
\usepackage{bm}
\usepackage{amsmath,amssymb}
\usepackage{color,multirow}

\usepackage{graphicx}

\begin{document}

\title{Majorana edge states and topological 
properties in 1D/2D Rashba semiconductor proximity coupled to iron-based 
superconductor }

\author{Hiromi Ebisu$^1$, Keiji Yada$^2$, Hideaki Kasai$^{1,3}$, and Yukio Tanaka$^2$}
\inst{$^1$Department of Applied Physics, Osaka University, 2-1 Yamadaoka, Suita, Osaka 565-0871, Japan\\
$^2$Department of Applied Physics, Nagoya University, Nagoya 464-8603, Japan\\
$^3$Center for Atomic and Molecular Technologies, Osaka University, 2-1 Yamadaoka, Suita, Osaka 565-0871, Japan
}





\abst{
 We study Majorana edge states 
and their topological properties 
of one-dimensional(1D) and two-dimensional(2D) Rashba semiconductor 
deposited on iron-based superconductor 
under the applied Zeeman field for various directions. 
Using the recursive Green's function method,
we calculate the local density of states(LDOS) both for 
$s_{\pm}$ and $s_{++}$-wave pairings. 
We elucidate that it shows  anisotropic response to the applied 
Zeeman field specific to Majorana edge states. 
This anisotropy can be understood by 
the winding number, which shows  
whether the present system is topological or not. 
The resulting LDOS and winding numbers  for $s_{\pm}$ and $s_{++}$-wave pairings are significantly different at the lower Zeeman field. 
These results serve as a guide to determine the pairing symmetry of 
iron-pnictide. 
}


\maketitle
\thispagestyle{empty}

\section{Introduction}
Topological superconducting systems with 
gapless surface Andreev bound states (SABSs) 
are focused in condensed matter physics now. 
\cite{Read,qi11,alicea12,tanaka12}
In these systems, Majorana fermions  
\cite{wilczek09,Ivanov01,Read,alicea12}
appear as SABSs or vortex core states. 
For the future application of a fault tolerant quantum computation, 
Majorana fermions are important ingredients  
\cite{nayak} since they obey non-Abelian statistics. 
Topological superconductivity with Majorana fermions 
was originally discussed in 
spinless triplet $p$-wave superconductors.\cite{Read,Ivanov01,Kitaev01}
However, to realize spinless triplet superconductor is not easy in 
real solid state systems. \par

Fu and Kane proposed that topological superconductivity 
is possible in ferromagnet  / spin-singlet $s$-wave 
superconductor hybrid system deposited on the surface of topological insulator.\cite{Fu08}
To elucidate the physical properties  of Majorana fermions in this system
has become a hot topic and 
several theoretical \cite{Fu09,Akhmerov09,Tanaka09,Linder10} 
and experimental researches \cite{veldhorst12,williams12} have been presented. 
The key concept for the generation of topological superconductivity from
conventional spin-singlet $s$-wave superconductor is the 
simultaneous existence of 
spin-orbit coupling and the broken time reversal symmetry 
like Zeeman field.\cite{Fujimoto09,STF09}
In the presence of the Rashba spin-orbit coupling, the energy dispersion of 
free electrons splits into two.
Further, by introducing Zeeman field and 
tuning chemical potential, one of the spin-polarized Fermi surface disappears. 
Then, exotic spinless metallic state, i.e. helical metallic state, 
is attainable where the direction of the spin of an electron is 
rocked to the momentum. \par
A semiconductor quantum well coupled to 
an $s$-wave superconductor and a ferromagnetic insulator is a possible 
candidate,\cite{Jay,Alicea10} where one-dimensional 
chiral Majorana edge state is generated as a SABS. 
Also Majorana edge state was proposed 
in semiconductor nano-wires deposited on the surface of 
spin-singlet $s$-wave superconductor in the presence of 
external Zeeman field. \cite{oreg10,lutchyn10}
Since Majorana edge state is generated as the end state of 
the nano-wire, braiding operation might be possible by   
making network of wires.\cite{alicea11}
It is noted that several experiments supported  the 
existence of Majorana edge state 
\cite{Mourik,Deng,Das,Churchill,Finck,Lee}
through zero bias conductance peak
\cite{TK95,kashiwaya00,Bolech,law09,Wimmer,Beenakker13} 
or anomalous Josephson current.\cite{Rokhinson,Kwon}
\par
Besides these streams, there have been many researches  
about topological superconductivity in the time-reversal invariant (TRI) 
systems without the Zeeman field. 
\cite{Schnyder08,TYBN08,Kitaev09,Fu10,TMYYS10,STYY11, 
Qi09,Qi2010,Msato,Teo,Ortiz,Nakosai1,Nakosai2}
These systems belong to so called DIII class in the periodic 
table \cite{Schnyder08} of topological materials.
Doped topological insulator Bi$_{2}$Se$_{3}$ is one of the candidate materials 
of DIII class superconductor.\cite{hor10,sasaki11}
Although several theoretical and experimental works 
support the realization of  TRI topological superconductivity 
in this material, pairing symmetry and resulting 
Majorana modes are not fully determined yet.  

There are several proposals to 
realize TRI topological superconductors (SCs) based on 
hybrid systems proximity coupled to 
unconventional superconductors.\cite{Nakosai2,Wong,Nayak12}
Recently, Zhang, Kane and Mele proposed 
hybrid systems 
combining doped Rashba semiconductors (RSs) and iron-based SCs.\cite{FanZhang}
They have assumed that  spin-singlet $s_{\pm}$-wave 
pairing is realized in iron-based 
SCs.\cite{Ishida2009,Mazin2008,Kuroki2008}
Then, the resulting pair potential in semiconductor 
has a sign change between $\Gamma$ and $M$ points. 
Kramers pair of Majorana edge states is generated 
at the boundary of the 2D (1D) TRI hybrid 
superconducting system due to the sign change of the pair potential 
between the two Fermi surfaces. 
However, the pairing symmetry of the iron based superconductors is 
still in the hot debate. 
So called $s_{++}$-wave pairing with no sign change of the pair potential 
is also possible by the orbital fluctuation.\cite{Kontani,Onari2012,Yanagi}
Although, tunneling spectroscopy of iron-based superconductors 
have been calculated,\cite{Golubov2009,Yada2013} the significant 
qualitative difference of the line shape of the tunneling conductance 
between $s_{\pm}$-wave and $s_{++}$-wave pairing does not exist. 
The tunneling conductance of normal metal / $s_{\pm}$-wave superconductor 
junctions is not distinct as compared to $d$-wave \cite{TK95,kashiwaya00} 
or $p$-wave. \cite{YTK97,YTK98,Tanuma2001,Tanuma2002}
superconductor junctions. 
Thus, it is highly encouraged to present a new idea to 
distinguish above two parings in a qualitative level. 

In the present paper, we study 
1D/2D edge states of doped Rashba semiconductor 
deposited on iron-based superconductors 
by applying the Zeeman field for various directions. 
Using recursive Green's function method, 
we calculate local density of states on the edge 
and the angle resolved local density of states in two-dimensional case 
both for $s_{\pm}$ and $s_{++}$-wave paring cases. 
We concentrate on the Majorana edge modes as an ABS and relevant 
local density of state (LDOS) including zero energy peak (ZEP). 
We elucidate that the anisotropic response to the Zeeman field 
stems from the mirror reflection symmetry. 
\cite{Zhang13,Ueno,Yao,Morimoto,Sato2014}
We also calculate winding numbers of the Hamiltonian and 
analyze topological properties. 
The resulting LDOS and winding numbers  
are seriously different between these two pairing cases. 
It can be concluded that Rashba semiconductor / iron-based superconductor 
hybrid junctions are useful to determine the pairing symmetry of iron-pnictide. \par

The organization of this paper is as follows. 
In section \ref{sec2}, we explain the model, Hamiltonian, and the formulation. 
In section \ref{sec3a}, we study a 1D Rashba semiconductor nanowire / 
iron-based superconductor 
hybrid system. We calculate energy spectrum of the bulk nanowire and  
LDOS on the edge for various directions of the applied Zeeman field. 
In section \ref{sec3b}, we study a 2D Rashba semiconductor layer / iron-based superconductor hybrid system. 
Angle resolved LDOS for fixed momentum $k_{y}$ parallel to the surface 
and angle averaged LDOS on the edge are calculated. 
Both in 3(a) and 3(b), we compare the results of $s_{\pm}$-wave and 
$s_{++}$-wave case. 
In section \ref{sec4}, we interpret calculated results in section \ref{sec3} based on winding number of the system. 
We also discuss   the reason why LDOS is sensitive to the 
direction of the applied Zeeman field 
in terms of the Ising like spin of Majorana edge sate. 
In section \ref{sec5}, we summarize our results.

\section{Formulation}\label{sec2}
\label{Formulation}
\begin{figure}
\begin{center}
\includegraphics[width=9cm]{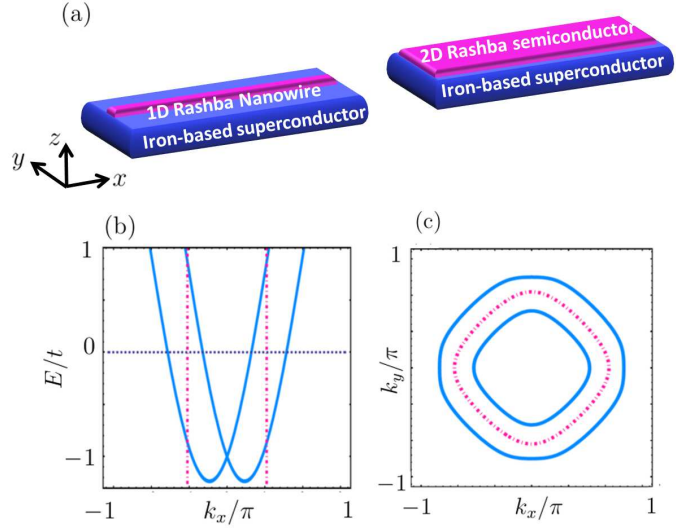}
\caption{(a)(left)A 1D Rashba semiconductor nanowire(left) 
and a 2D Rashba semiconductor layer(right) deposited on iron-based superconductors.(b)
The energy dispersion of a nanowire in the normal state for $t$=1, $\lambda_R=0.5$ (solid line)
and  the position of Fermi energy $\mu=-1$ (dotted line). 
Dot-dashed line denotes nodal lines  for $\Delta_0=-0.2$, $\Delta_1=0.2$. 
The sign of the pair potential inside(outside) nodal lines is 
positive(negative). 
(c)Fermi surface 
of a Rashba layer for $t$=1, $\lambda_R=0.5$, and $\mu=-1$ (solid line). 
Dot-dashed line denotes nodal line for $\Delta_0=-0.2$, $\Delta_1=0.2$. }
\label{Fermi}
\end{center}
\end{figure}

In this section, we introduce model Hamiltonian of 1D and 2D Rashba 
semiconductors 
deposited on iron-based  superconductors 
as shown in Fig. \ref{Fermi}(a) and 
write a formula of recursive Green's function to calculate LDOS 
on the edge.

\subsection{Model}\label{sec2a}

Model Hamiltonian of 1D Rashba doped semiconductor nanowire 
in momentum space reads 
\begin{eqnarray}
\mathcal{H}_0^{\text{1D}}(k_x)&=&(-2t\cos k_x-\mu)\sigma_0\tau_z\notag
+ 2\lambda_R\sin k_x\sigma_y\tau_z\notag\\
&&-\Bigl(\Delta_0+2\Delta_1\cos k_x\Bigr)\sigma_y\tau_y,
\end{eqnarray}
where $t$ is the nearest neighbor hopping, 
$\mu$ is a chemical potential, 
$\lambda_R$ is Rashba spin-orbit coupling. 
We have chosen the value of chemical potential $\mu$, so that 
charge transport properties of 
Rashba semiconductor becomes metallic by doping effect. 
$\Delta_0$ and $\Delta_1$ are proximity induced 
pair potentials in 1D Rashba semiconductor. 
In general, these values are less than that in bulk 
iron-based superconductor. 
$\bf{\sigma}$ and $\mathbf{\tau}$ are the Pauli matrices in spin space and 
electron-hole space, respectively.  
When $|\mu-t\Delta_0/\Delta_1|<2\lambda_R\sqrt{1-\Delta_0^2/(4\Delta_1^2)}$, 
quasiparticle feels different sign of pair potential 
at the inner and the outer Fermi points.  
Then, we verify that $s_{\pm}$-wave pairing is realized in the 1D Rashba semiconductor. \par
The corresponding 
model Hamiltonian of 2D Rashba semiconductor in momentum space is given by
\begin{eqnarray}
\mathcal{H}^{\text{2D}}_0(k_x,k_y)&=&\Bigl( -2t(\cos k_x+\cos k_y)-\mu\Bigr)\sigma_0\tau_z\notag\\
&&-2\lambda_R\sin k_y\sigma_x\tau_0+ 2\lambda_R\sin k_x\sigma_y\tau_z\notag\\
&&-\Bigl(\Delta_0+2\Delta_1(\cos k_x+\cos k_y)\Bigr)\sigma_y\tau_y
\end{eqnarray}      
When we choose 
$|\mu-t\Delta_0/\Delta_1|<2\lambda_R\sqrt{2-\Delta_0^2/(8\Delta_1^2)}$, 
$s_{\pm}$-wave pairing is realized.\par
The nodal lines for $s_{\pm}$-wave case in 
1D and 2D are shown in 
Figs. 1(b) and (c), respectively. 
For $s_{\pm}$-wave pairing, 
Majorana Kramers Doublets emerge as edge states, 
and DIII class topological superconducting state is realized.\cite{FanZhang}
The applied Zeeman field is given by 
\begin{equation}
\hat{V}_{i}=\begin{cases}
    V_x\sigma_x\tau_z \ (i=x, \; x\text{-direction})\\
    V_y\sigma_y\tau_0 \ (i=y, \; y\text{-direction})\\
   V_z\sigma_z\tau_z \ (i=z, \; z\text{-direction}).
  \end{cases}
\end{equation}
Thus, the total Hamiltonian in 1D and 2D with the Zeeman field is given by 
\begin{equation}
\mathcal{H}^{\text{1D(2D)}}_{i}(k_x (k_x,k_y))=\mathcal{H}_0^{\text{1D(2D)}}(k_x (k_x,k_y))
+\hat{V}_{i}.
\end{equation}  
In the presence of the Zeeman field, the time-reversal symmetry 
is broken and topological nature of the Majorana edge states is 
changing.

\subsection{Recursive Green's function}\label{sec2b}

\begin{figure}
\begin{center}
\includegraphics[width=9cm]{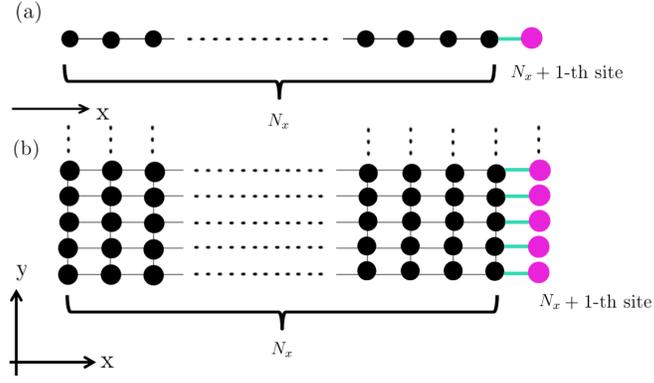}
\caption{Schematic pictures of the process of 
calculating recursive Green's function in (a)1D case and  (b)2D case.}
\label{green}
\end{center}
\end{figure}

First, we derive a general formula of the recursive Green's function in 1D. 
Consider a system which has finite length in the $x$-direction. 
Suppose the number of the sites is $N_x$, and add one more site in the 
$x$-direction (Fig. \ref{green}(a)).
Based on the Dyson equation, we obtain 
\begin{equation}
\label{dyson}
G=G_0+G_0TG
\end{equation}
for Green's function $G$ of this system, where
\begin{eqnarray}
G_0&=&\left| N_x \right\rangle G_{N_x,N_x}\left\langle N_x\right|+\left| N_x+1 \right\rangle g_{\text{iso}}\left\langle N_x+1\right|\notag\\
T&=&\left| N_x+1 \right\rangle T_{N_x+1,N_x}\left\langle N_x\right|+h.c.
\end{eqnarray}
$G_{N_x,N_x}$ is the Green's function at $N_x$-site, 
$g_{iso}$ is the isolated 
Green's function at $N_x+1$-site, and $T_{N_x+1,N_x(N_x,N_x+1)}$ is a 
transfer integral from $N_x$-th to $N_x+1$-th ($N_x+1$ to $N_x$) site. 
We derive following two equations from 
Eq. \eqref{dyson}, 
\begin{eqnarray}
G_{N_x+1,N_x+1}&=&g_{\text{iso}}+g_{\text{iso}}T_{N_x+1,N_x}G_{N_x,N_x+1}\label{n1n1}\\
G_{N_x,N_x+1}&=&G_{N_x,N_x}T_{N_x,N_x+1}G_{N_x+1,N_x+1}\label{nn1}.
\end{eqnarray}
By substituting  $G_{N_{x},N_{x}+1}$ in \eqref{n1n1} 
for \eqref{nn1}, we obtain 
\begin{eqnarray}
G_{N_x+1,N_x+1}&=&\Bigl(g^{-1}_{\text{iso}}-T_{N_x+1,N_x}G_{N_x,N_x}T_{N_x,N_x+1}\Bigr)^{-1}
\end{eqnarray}
Once we know $G_{N_x,N_x}$, we obtain $G_{N_x+1,N_x+1}$.
In order to calculate the LDOS, we evaluate a retarded Green's function 
$G_{jj}(E + i0^{+})$ with site index $j$ and energy $E$ measured from the 
Fermi level, where  $0^{+}$ is an infinitesimal number. 
In the present model,  following relations 
\begin{eqnarray}
T_{j+1,j}=-t\sigma_0\tau-i\lambda_R\sigma_y\tau_z-\Delta_{1}\sigma_y\tau_y\;\;(j=1\sim N_x)\\
g_{\text{iso}}(E)=\Bigl(EI_{4\times4}- \mathcal{H}_{\text{iso}}\Bigr)^{-1}\\
 \mathcal{H}_{\text{iso}}= -\mu\sigma_0\tau_z-\Delta_0\sigma_y\tau_y+\hat{V}_{i}, 
\end{eqnarray}
are satisfied. 
After calculating $G_{1,1}(E)=g_{\text{iso}}(E)$, we can obtain 
Green's function of any number of sites in the $x$-direction. 
Finally, LODS on the edge is give by
\begin{equation}
\label{1d}
\rho^{\text{1D}}(E)=-\frac{1}{\pi}{\rm Im}\Bigr\{\text{Tr}[G_{N_x+1,N_x+1}(E+i0^{+})]\Bigl\},
\end{equation}
where $\rm{Im}$ denotes to choose imaginary part. 
In  the evaluation of the trace of $G_{N_x+1,N_x+1}(E+i0^{+})$, 
we sum up only the electronic part of $G_{N_x+1,N_x+1}(E+i0^{+})$. \par
In the 2D model, we assume that the boundary is flat 
as shown in Fig.\ref{green}(b). 
In this case, momentum $k_y$ is a good quantum number. 
Then, we can calculate the 
recursive Green's function for each $k_{y}$ in the similar way 
in 1D case. 
We obtain 
\begin{eqnarray}
T_{i+1,i}=-t\sigma_0\tau-i\lambda_R\sigma_y\tau_z-\Delta_{1}\sigma_y\tau_y\;\;(i=1\sim N)\\
g_{\text{iso}}(E,k_y)=\Bigl(EI_{4\times4}- \mathcal{H}_{\text{iso}}(k_y)\Bigr)^{-1}\\
 \mathcal{H}_{\text{iso}}(k_y)=  \Bigl(-2t(\cos k_x+\cos k_y)-\mu\Bigl)\sigma_0\tau_z\notag\\
-2\lambda_R\sin k_y\sigma_x\tau_0\notag\\
-\Bigl(\Delta_0+2\Delta_1\cos k_y\Bigr)\sigma_y\tau_y+\hat{V}_{i}.
\end{eqnarray}
Angle resolved local density of states (ARLDOS) for each $k_{y}$ is given by 
\begin{equation}
\label{2d1}
D^{\text{2D}}(E,k_y)=-\frac{1}{\pi}{\rm Im}\Bigr\{\text{Tr}[G_{N_{x}+1,N_{x}+1}(E+i0^{+},k_y)]\Bigl\}.
\end{equation}
To obtain the LDOS, we calculate
\begin{equation}
\label{2d2}
\rho^{\text{2D}}(E)=\frac{1}{N_y}\sum_{k_y}D^{\text{2D}}(E,k_y),
\end{equation}
where, $N_y$ is a number of sites in the $y$-direction.  
Using Eqs. \eqref{1d}, \eqref{2d1}, 
and \eqref{2d2}, we obtain LDOS on the edge 
of 1D model and  ARLDOS and LDOS in the 2D model in 
the presence of the Zeeman field ($x$, $y$, $z$-directions) 
both for $s_{\pm}$ and $s_{++}$-wave pairing.

\section{Calculated results of LDOS}\label{sec3}
In this section, we show the results of LDOS in 1D and ARLDOS and LDOS in 2D. 
In this paper, we fix parameters as $t$=1, $\lambda_R=0.5$, $\mu=-1$. 
%
We choose 
$\Delta_0=-0.2$ and  $\Delta_1=0.2$ for $s_{\pm}$-wave pairing 
and $\Delta_0=0.2$ and $\Delta_1=0$ for $s_{++}$-wave pairing.

\subsection{1D Rashba semiconductor}\label{sec3a}
First, we show the LDOS on the edge of a 1D Rashba semiconductor nanowire 
deposited on iron-based superconductors with the Zeeman field in 
the $x$, $y$, and $z$-directions 
for $s_{\pm}$-wave case (Fig. 3).
Due to the presence of
Majorana Kramers doublets realized without the Zeeman field, 
we can clearly see the zero energy peak of LDOS of edge state for 
$V_{i}=0$ ($i=x,y,z$) (Fig. 3). 
Since this Kramers doublet is topologically protected by the time-reversal 
symmetry, 
a pair of Kramers doublet is lifted by the applied Zeeman field in general.
Here, we show the LDOS under the applied Zeeman field in the $x$, $y$ and $z$-direction in Figs.3 (a), (b) and (c), respectively.
A zero energy peak in the LDOS remains 
in the presence of the Zeeman field in 
the $x$ and $z$-direction 
while it disappears for the $y$-direction.
This suggests that remaining symmetry of the Hamiltonian, which is discussed below, protect the Majorana edge  states under the Zeeman field in 
$x$ and $z$-directions. 
Next, we see LDOS at the edge under the Zeeman field in the $x$-direction.
We can also see a closing of the bulk energy gap at $V_x/t\sim 1$ and $3$.
A zero energy peak survives after the first gap closing at $V_x/t \sim 1$ 
as shown in Fig. 3(d).
However, this zero energy peak disappears after the second gap closing at $V_x/t\sim 3$. 
The similar features are obtained when the Zeeman field 
is along $z$-direction. 
\par
In the case of $s_{++}$-wave pairing, 
no zero energy state (ZES), i.e., Majorana edge state,  
appears on the edge without the Zeeman field, 
but if we apply  the Zeeman field in $x$, $z$-directions, 
LDOS has a ZEP  at $V_{x,(z)}/t \sim 1$ (Fig. 4 (a)(c)) 
\cite{oreg10,lutchyn10} 
which is quite different from $s_{\pm}$-wave pairing case. 
When, we apply the Zeeman field in the $y$-direction, 
ZEP is not seen at all (Fig. 4(b)). 
These features are essentially the same with 
those in conventional $s$-wave superconductor / 1D Rashba semiconductor 
hybrid systems. 
For both $s_{\pm}$ and $s_{++}$cases, LDOS is sensitive to the 
direction of the Zeeman field. 
In the presence of the sufficient large magnitude of $V_{i}$ ($i=x,y,z$), 
LDOS has a ZES for $i=x$ and  $i=z$ but not for $i=y$.
\begin{figure}
\begin{center}
\includegraphics[width=9cm]{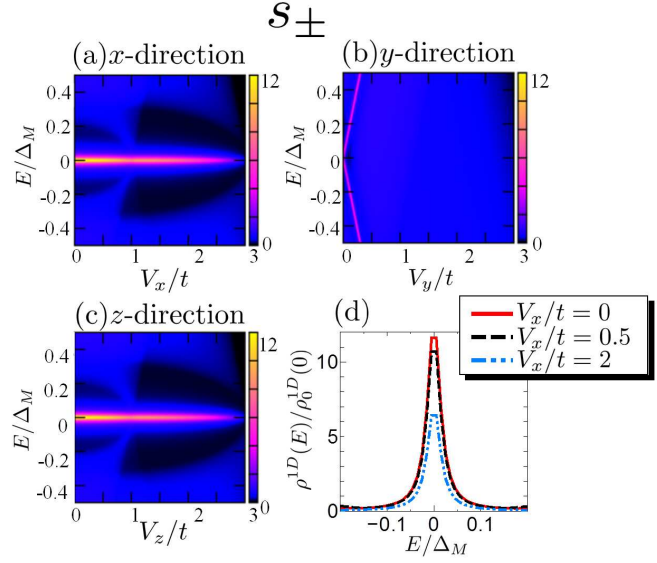}
\caption{(a)$\sim$(c)Intensity plot of LDOS of a 1D Rashba semiconductor nanowire on $s_{\pm}$-wave superconductor with the Zeeman field in (a)$x$,(b) $y$, and (c)$z$-direction. 
(d)Normalized LDOS of a Rashba semiconductor nanowire on  $s_{\pm}$-wave superconductor under the Zeeman field in the $x$-direction, for $V_x/t=0$(solid line),$V_x/t=0.5$(dashed line), $V_x/t=2$(dot-dashed line). 
LDOS is normalized by its value in the normal state 
at $E=0$, and the energy is normalized by $\Delta_M=\Delta_0+2\Delta_1$. }

\end{center}
\end{figure}

\begin{figure}
\begin{center}
\includegraphics[width=9cm]{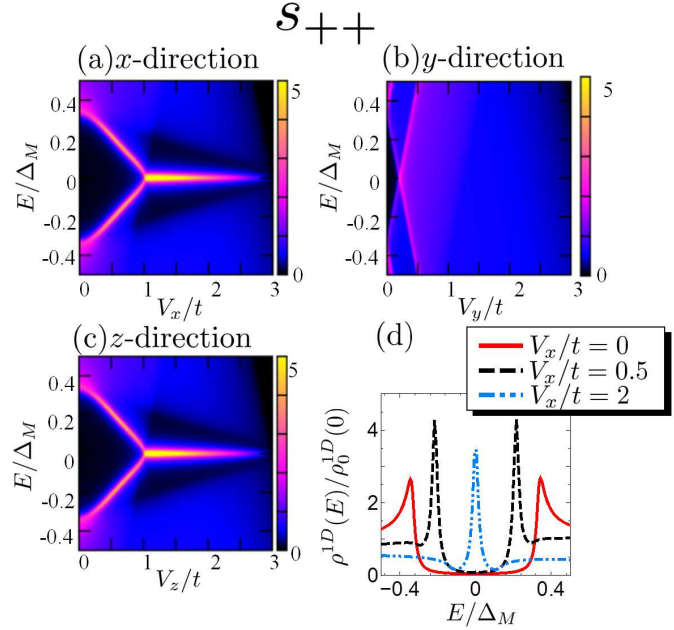}
\caption{
(a)$\sim$(c)Intensity plot of LDOS of a 1D Rashba semiconductor nanowire on $s_{++}$-wave superconductor with the Zeeman field in (a)$x$, (b)$y$, and (c)$z$-direction.(d)Normalized LDOS of a Rashba semiconductor nanowire on  $s_{\pm}$-wave superconductor under the Zeeman field in the $x$-direction, for $V_x/t=0$(solid line),$V_x/t=0.5$(dashed line), $V_x/t=2$(dot-dashed line). 
LDOS is normalized by its value in the normal state 
at $E=0$, and the energy is normalized by $\Delta_M=\Delta_0+2\Delta_1$.}

\end{center}
\end{figure}

\subsection{2D Rashba semiconductor}\label{sec3b}
\begin{figure}
\begin{center}
\includegraphics[width=9cm]{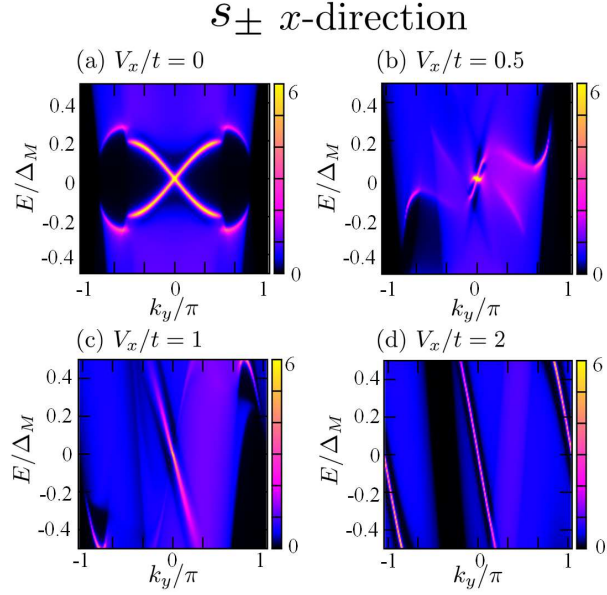}
\caption{ARLDOS of a 2D Rashba semiconductor layer in the case of $s_{\pm}$-wave pairing under the Zeeman field in the $x$-direction for $V_x/t=$(a)0, (b)0.5, (c)1, and (d)2. ARLDOS is normalized by its value in the normal state 
 at $k_y=0$ and  $E=0$. Energy is normalized by $\Delta_M=\Delta_0+4\Delta_1$.}
\label{x2D_s+-}
\end{center}
\end{figure}

\begin{figure}
\begin{center}
\includegraphics[width=9cm]{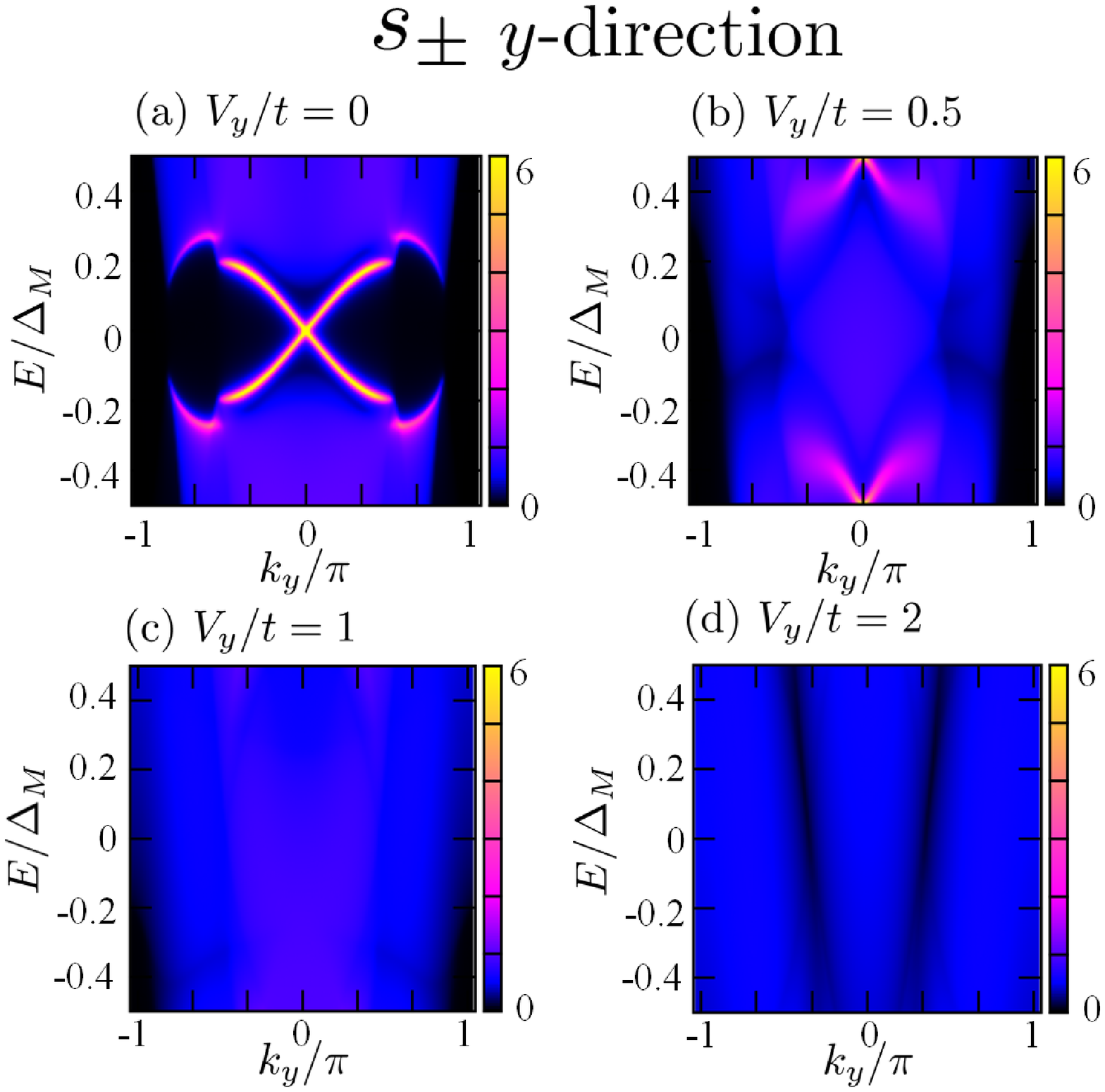}
\caption{ARLDOS of a 2D Rashba semiconductor layer in the case of $s_{\pm}$-wave pairing under the Zeeman field in the $y$-direction for $V_y/t=$(a)0, (b)0.5, (c)1, and (d)2. ARLDOS is normalized by its value in the normal state 
 at $k_y=0$ and  $E=0$. Energy is normalized by $\Delta_M=\Delta_0+4\Delta_1$.}
\label{y2D_s+-}
\end{center}
\end{figure}

\begin{figure}
\begin{center}
\includegraphics[width=9cm]{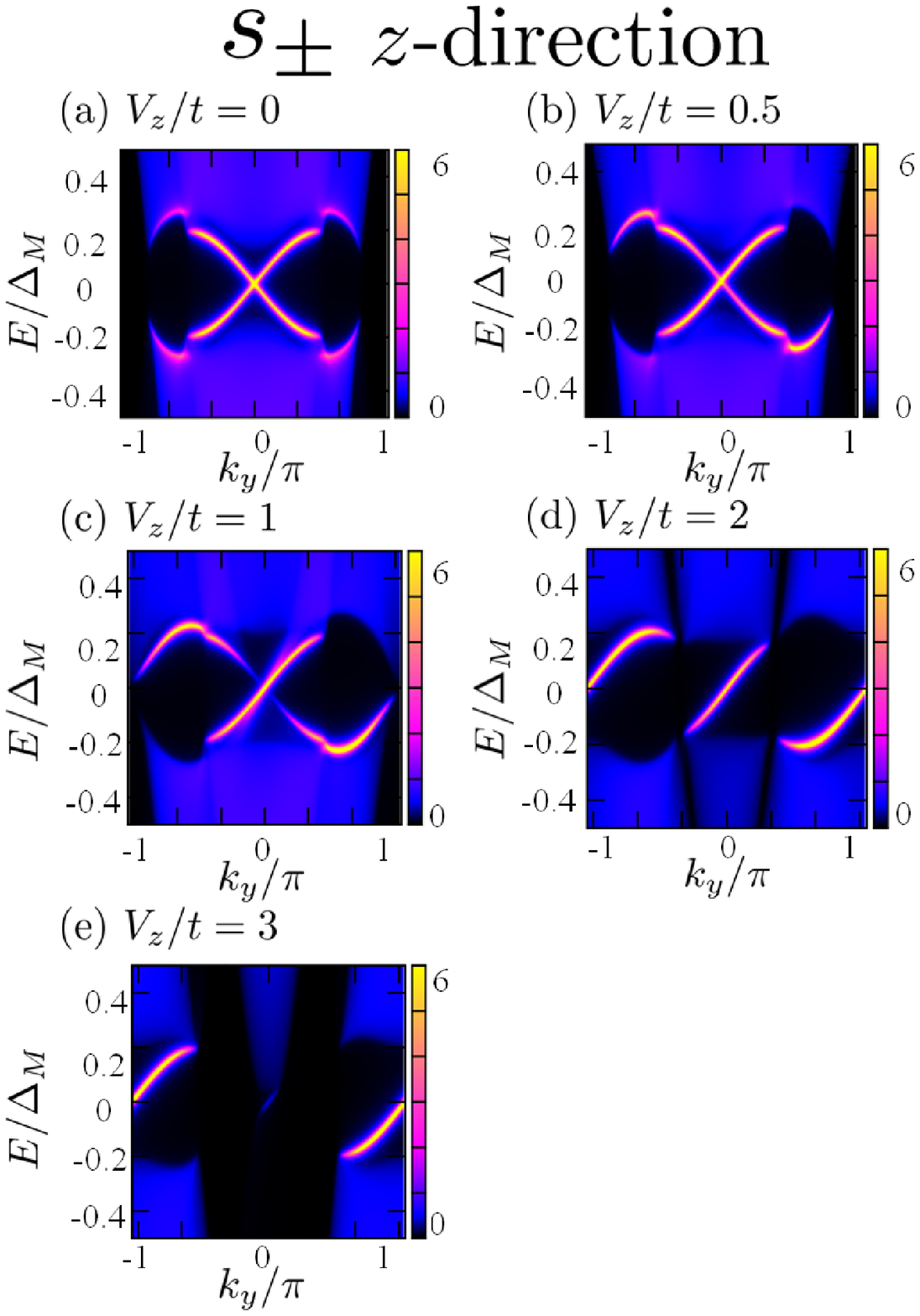}
\caption{ARLDOS of a 2D Rashba semiconductor layer in the case of $s_{\pm}$-wave pairing under the Zeeman field in the $z$-direction for $V_z/t=$(a)0, (b)0.5, (c)1, (d)2, and (e)3. ARLDOS is normalized by its value in the normal state 
 at $k_y=0$ and  $E=0$. Energy is normalized by $\Delta_M=\Delta_0+4\Delta_1$.}
\label{2D_s+-}
\end{center}
\end{figure}

\begin{figure}
\begin{center}
\includegraphics[width=9cm]{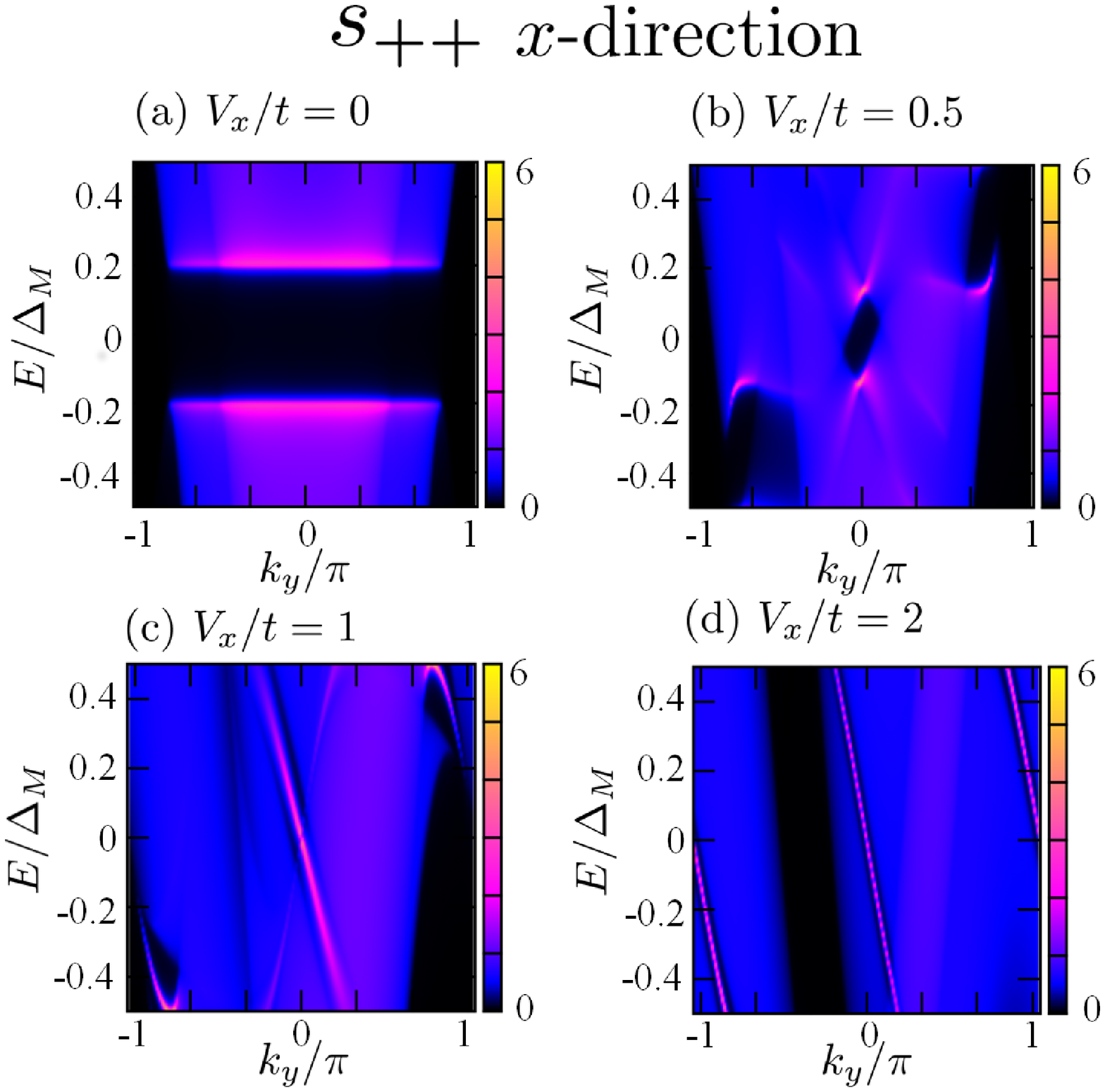}
\caption{ARLDOS of a 2D Rashba semiconductor layer in the case of $s_{++}$-wave pairing under the Zeeman field in the $x$-direction for $V_x/t=$(a)0, (b)0.5, (c)1, and (d)2. ARLDOS is normalized by its value in the normal state 
 at $k_y=0$ and  $E=0$. Energy is normalized by $\Delta_M=\Delta_0+4\Delta_1$.}
\label{x2D_s++}
\end{center}
\end{figure}

\begin{figure}
\begin{center}
\includegraphics[width=9cm]{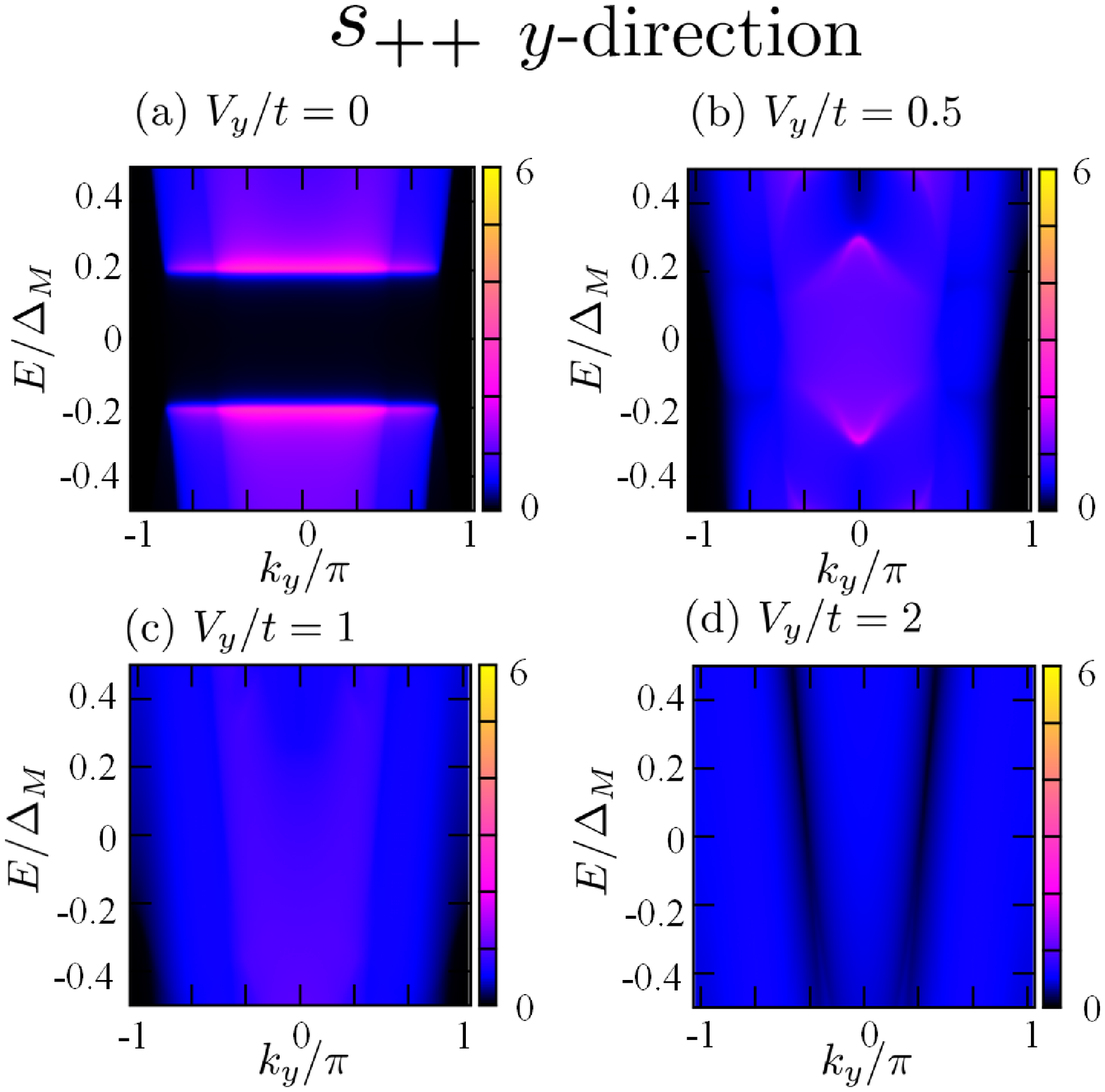}
\caption{ARLDOS of a 2D Rashba semiconductor layer in the case of $s_{++}$-wave pairing under the Zeeman field in the $y$-direction for $V_y/t=$(a)0, (b)0.5, (c)1, and (d)2. ARLDOS is normalized by its value in the normal state 
 at $k_y=0$ and  $E=0$. Energy is normalized by $\Delta_M=\Delta_0+4\Delta_1$.}
\label{y2D_s++}
\end{center}
\end{figure}

\begin{figure}
\begin{center}
\includegraphics[width=9cm]{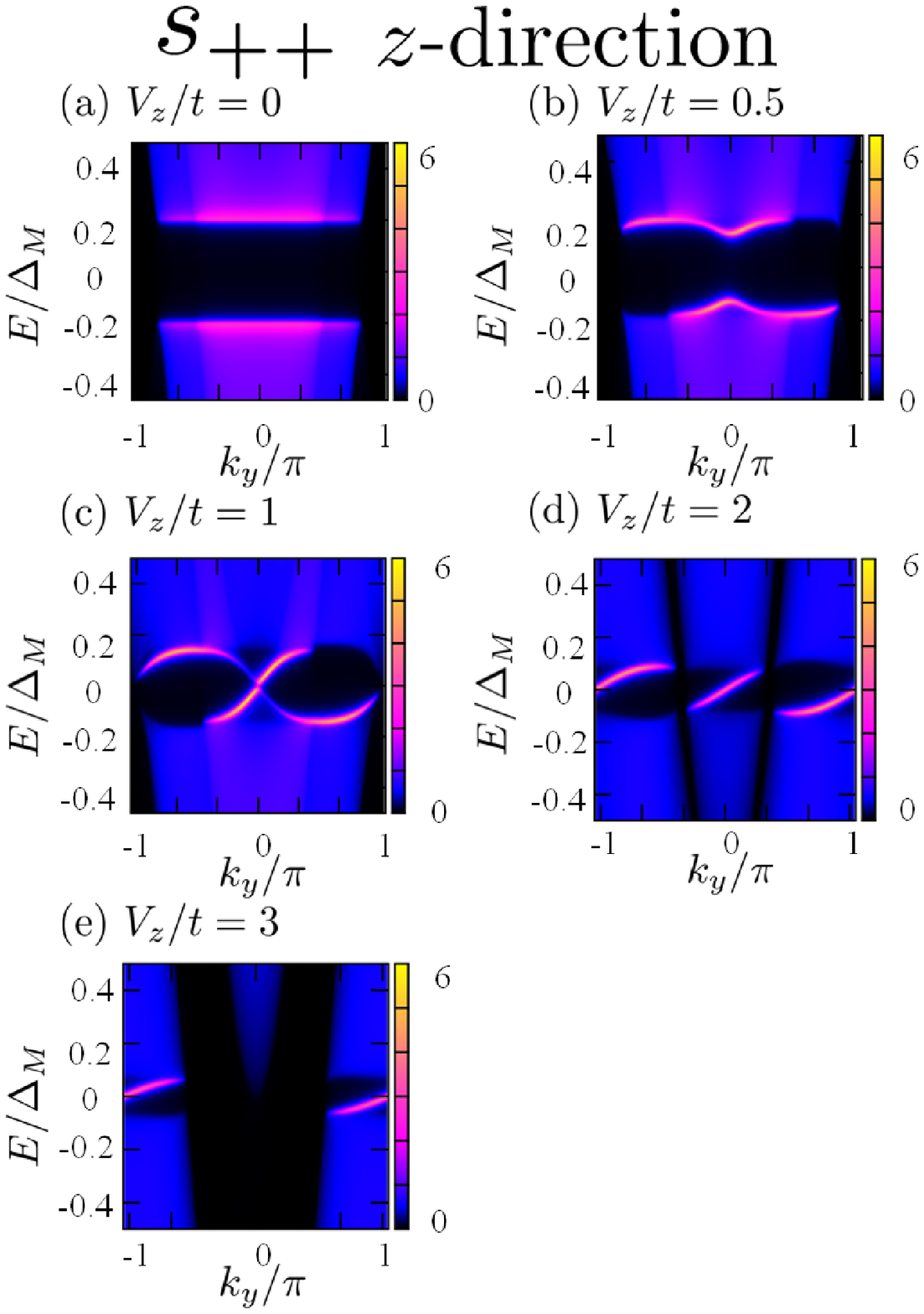}
\caption{ARLDOS of a 2D Rashba semiconductor layer in the case of $s_{++}$-wave 
pairing under the Zeeman field in the $z$-direction for $V_z/t=$(a)0, (b)0.5, (c)1, (d)2, and (e)3. ARLDOS is normalized by its value in the normal state 
 at $k_y=0$ and  $E=0$. Energy is normalized by $\Delta_M=\Delta_0+4\Delta_1$.}
\label{2D_s++}
\end{center}
\end{figure}

\begin{figure}
\begin{center}
\includegraphics[width=9cm]{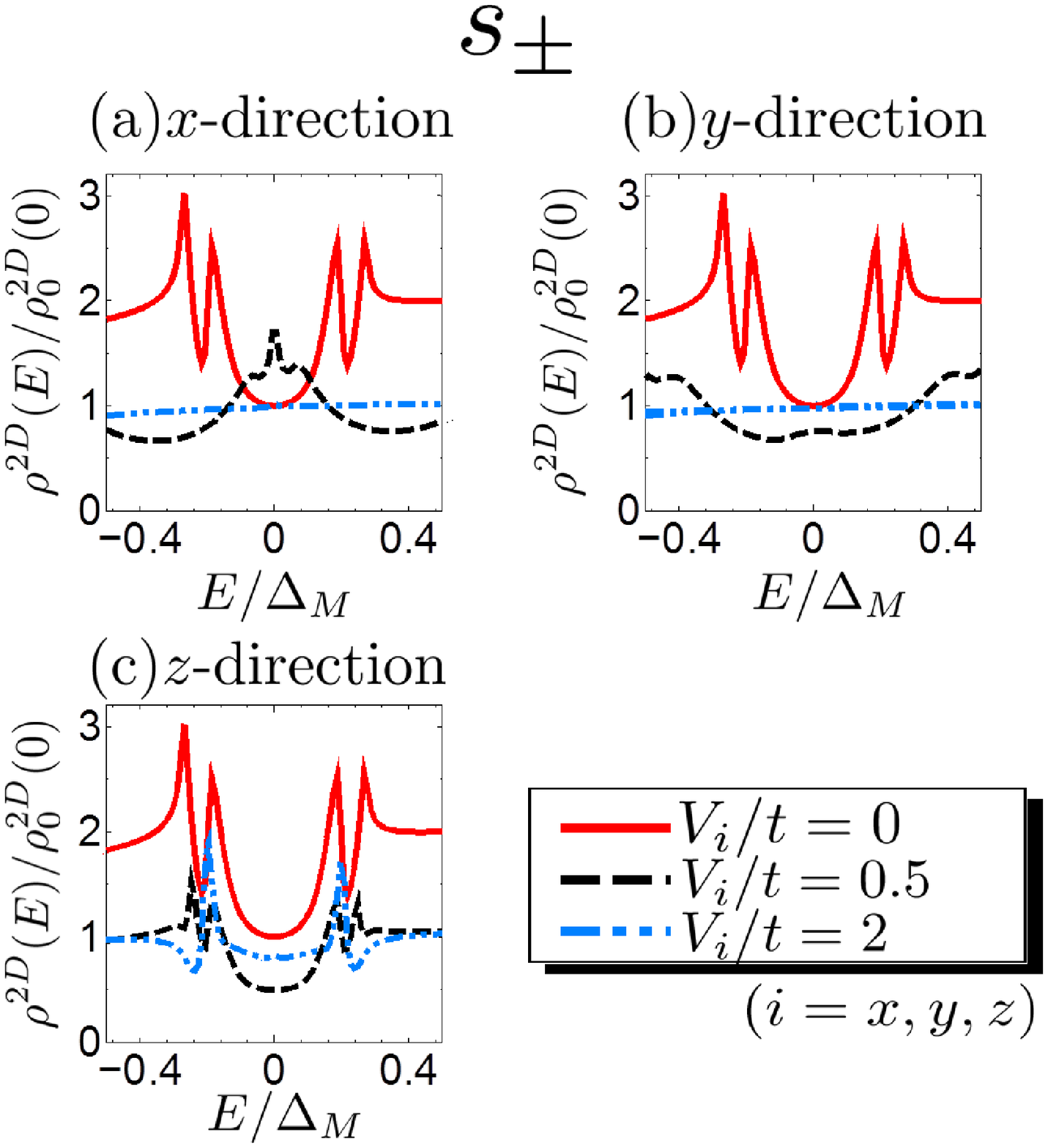}
\caption{(a)(b)(c)Normalized LDOS of a 2D Rashba semiconductor layer on $s_{\pm}$-wave superconductor with the Zeeman field $V_{i}/t=0$ (solid line),  $V_{i}/t=0.5$ (dashed line), and $V_{i}/t=2$ (dot-dashed line) in (a)$x$-direction, (b)$y$-direction, and (c)$z$-direction respectively. LDOS is normalized by its values in the normal state at $E=0$.}
\label{l2D_s+-}
\end{center}
\end{figure}

\begin{figure}
\begin{center}
\includegraphics[width=9cm]{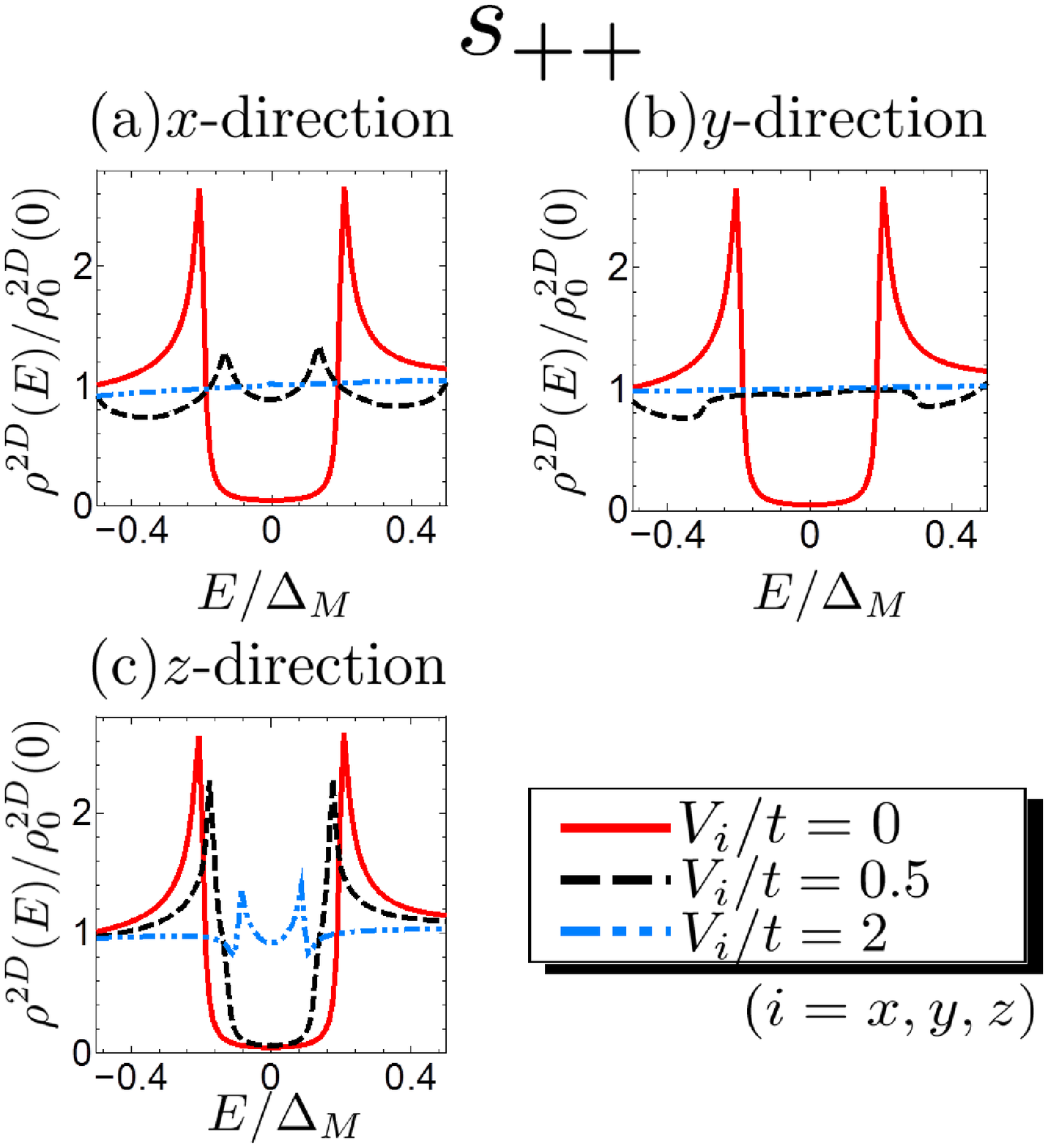}
\caption{(a)(b)(c)Normalized LDOS of a 2D Rashba semiconductor layer on $s_{++}$-wave superconductor with the Zeeman field $V_{i}/t=0$ (solid line),  $V_{i}/t=0.5$ (dashed line), and $V_{i}/t=2$ (dot-dashed line) in (a)$x$-direction, (b)$y$-direction, 
(c)$z$-direction, and  respectively. LDOS is normalized by its values in the normal state at $E=0$.}
\label{l2D_s++}
\end{center}
\end{figure}

ARLDOS of 2D Rashba semiconductor on 
iron-based superconductor 
is shown in Figs. \ref{x2D_s+-} to \ref{2D_s++} 
both for $s_{\pm}$ and $s_{++}$-wave pairing cases. 
The resulting LDOS of 
$s_{\pm}$-wave and $s_{++}$-wave are shown in Figs. \ref{l2D_s+-} and 
\ref{l2D_s++}, respectively. Without the Zeeman field, helical edge modes 
crossing at $k_{y}=0$  appear in  $s_{\pm}$-wave pairing
(Figs. \ref{x2D_s+-}(a),\ref{y2D_s+-}(a),\ref{2D_s+-}(a)),
while it does not in $s_{++}$-wave case 
(Figs. \ref{x2D_s++}(a),\ref{y2D_s++}(a),\ref{2D_s++}(a)).  
The resulting angular averaged LDOS $\rho^{2D}(E)$ 
at $E=0$ has a  nonzero value 
for $s_{\pm}$-wave pairing(solid lines in Fig. 11) 
while it becomes almost zero for $s_{++}$-wave pairing and has a typical 
U-shaped structure 
(solid lines in Fig. 12). 

If we apply the Zeeman field, ARLDOS changes seriously. 
In the case of $s_{\pm}$-wave pairing, 
ARLDOS is very sensitive to the direction of the Zeeman field. 
When the applied Zeeman field is along the $x$-direction, 
the helical modes once disappear and 
ARLDOS at $k_{y}=E=0$ is enhanced for $V_{x}=0.5t$ shown in 
Figs. \ref{x2D_s+-}(b). 
The resulting LDOS has a zero energy enhanced structure shown by dashed line 
in Fig. 11(a). With the further increase of $V_{x}$,
two chiral edge modes appear but are very close to continuum levels.
\cite{Wong2013}
Then, the resulting LDOS has an
almost flat curve without any structures as shown by dot-dashed line in 
Fig. 11(a). When the Zeeman field is applied in the  $y$-direction, helical edge mode 
disappears simultaneously 
with gap closing and there is no clear edge mode any more. 
The resulting LDOS is almost insensitive to $E$ 
as shown in Fig. 11(b). 
On the other hand, when  
the applied Zeeman field is along the $z$-direction,  
helical modes change into two chiral edge modes at $k_y=0$ and $k_y=\pi$ (Fig. \ref{2D_s+-}(d))  and then 
one of the chiral edge modes vanishes (Fig. \ref{2D_s+-}(e)). 
Reflecting on the change of ARLDOS, 
the lines shapes of LDOS have fine structures shown by 
dot-dashed and dashed lines in Fig. 11 (c).  \par

Next, we focus on $s_{++}$-wave pairing case where 
chiral Majorana edge mode is generated for 
the Zeeman field along the $x$ and $z$-directions.  
In this case, Majorana edge mode appears only after 
the applied Zeeman field exceeds a critical value.
\cite{Jay,Alicea10,oreg10,lutchyn10}
When the applied field is along the $x$-direction, 
although the edge mode appears, it is very near the continuum levels
(Fig. 8(d)) \cite{Wong2013}. 
The line shape 
of the resulting LDOS does not have a sufficient 
change  as a function of $E$ shown in 
(dot-dashed line in Fig. 12 (a)). 
On the other hand, when the Zeeman field is applied 
in the $y$-direction, the bulk energy gap closes 
without any  generation of the edge mode shown in Fig. 9. 
For the $z$-direction case, 
two chiral edge modes are generated at  
$k_y=0$ and $k_y=\pi$ (Fig. \ref{2D_s++}(d))  and then 
one of the chiral edge mode vanishes (Fig. \ref{2D_s++}(e)). 
These features are similar to those in $s_{\pm}$-wave pairing case 
(Fig. \ref{2D_s+-}). \par
To summarize, the resulting ARLDOS and LDOS are sensitive to the 
direction of the applied Zeeman field both for $s_{++}$-wave 
and $s_{\pm}$-wave pairings. 
The difference between $s_{\pm}$-wave and $s_{++}$-wave pairings 
becomes clear when the applied  Zeeman field is along the $z$-direction 
with  $0 < V_z/t < 1$ as seen from Figs. 7 and 10.

\begin{figure}
\begin{center}
\includegraphics[width=9cm]{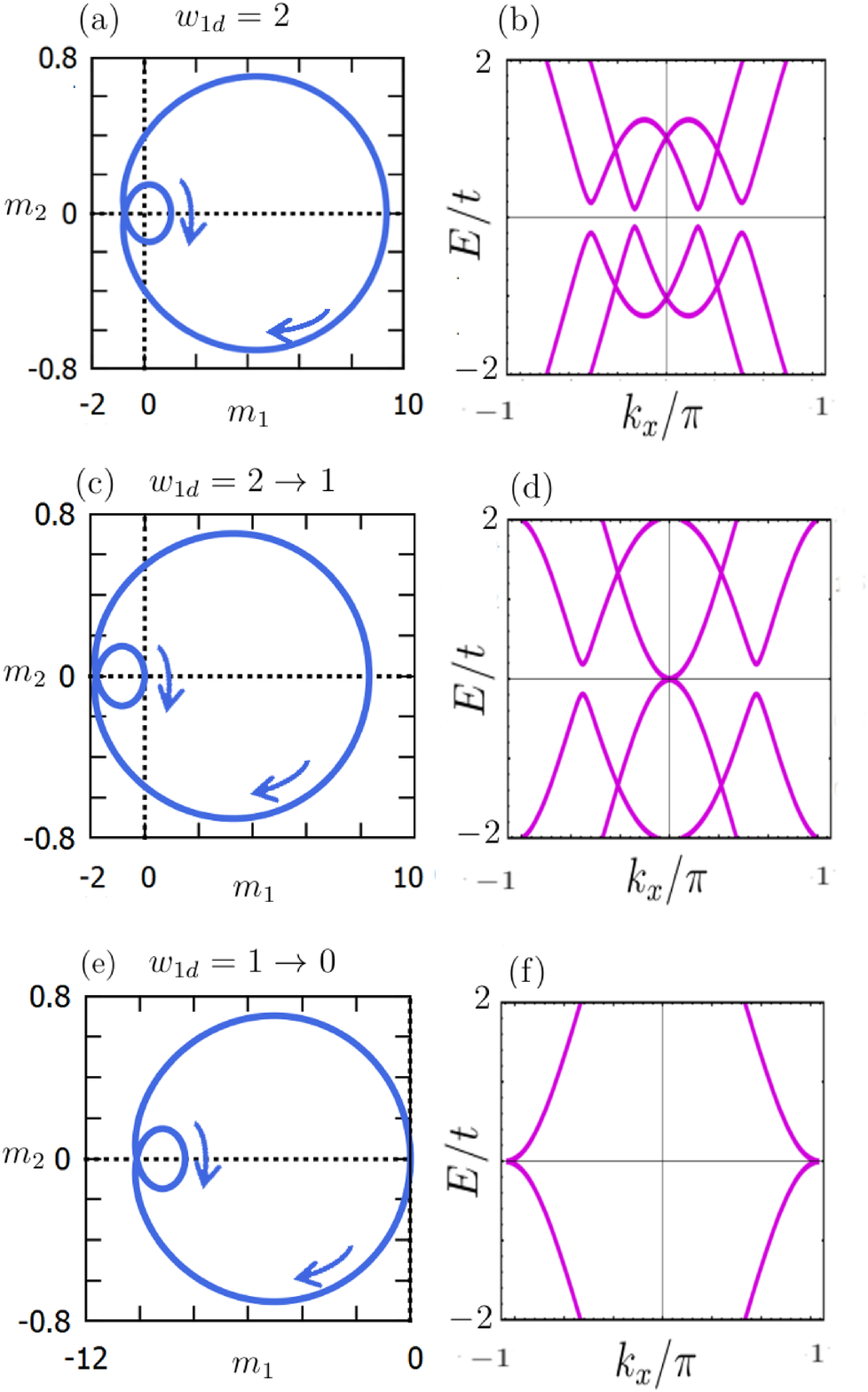}
\caption{(left)Trajectories of det$q(k_x)$  and (right) energy dispersion  of  $\mathcal{H}^{\text{1D}}_x(k_x)$ (a)(b) without the Zeeman field, (c)(d) with the Zeeman field in the $x$-direction $V_x/t=1.02$, and (e)(f)$V_x/t=3.06$, respectively. $m_1$ denotes real part of det$q(k_x)$ and $m_2$ is imaginary part of det$q(k_x)$. }
\label{winding}
\end{center}
\end{figure}

\section{Symmetry of the Hamiltonian and topological invariants}\label{sec4}
In this section, we elucidate that the anisotropic response of LDOS and LDOS 
to the Zeeman field stems from the mirror reflection symmetry in 1D by introducing 
winding number.
\subsection{1D case}\label{sec4a}
In 1D case, the Hamiltonian belongs to DIII class in periodic table 
\cite{Schnyder08} and has time-reversal symmetry and particle-hole symmetries.  
In addition these two, the system has a mirror reflection symmetry. 
These symmetries are given as follows: 
(i) time-reversal symmetry
\begin{equation}
\Theta^{\dagger}\mathcal{H}_0^{\text{1D}}(k_x)\Theta=(\mathcal{H}_0^{\text{1D}}(-k_x))^{*},\;\;\Theta=-i\sigma_y\tau_0,
\end{equation}
(ii)particle-hole symmetry 
\begin{equation}
C^{\dagger}\mathcal{H}_0^{\text{1D}}(k_x)C=
-(\mathcal{H}_0^{\text{1D}}(-k_x))^{*},\ \ C=\sigma_0\tau_x,
\end{equation}
(iii)mirror reflection symmetry (mirror plane is $xz$-plane)
\begin{equation}
\mathcal{M}^{\dagger}_{xz}\mathcal{H}_0^{\text{1D}}(k_x)\mathcal{M}_{xz}=\mathcal{H}_0^{\text{1D}}(k_x),\;\;\mathcal{M}_{xz}=i\sigma_y\tau_0.
\end{equation}
Combining three symmetries together, we can define operator $\Gamma^{1D}=\mathcal{M}_{xz}\Theta C$ which anti-commutes with $\mathcal{H}_0^{\text{1D}}(k_x)$.
\begin{equation}
\{\Gamma^{\rm 1D},\mathcal{H}_0^{\text{1D}}(k_x)\}=0.
\end{equation}
Now, we consider the effect of the Zeeman field term $\hat{V}_{i}$ in Eq. (3). 
When the Zeeman field is applied in the $x$, $z$-directions,
$\hat{V}_{i}$ anti-commutes with $\Gamma^{\text{1D}}$.
\begin{equation}
\label{anti}
\{\Gamma^{\rm 1D},\mathcal{H}_i^{\text{1D}}(k_x)\}=0, \;\;i=x,z.
\end{equation}
However, $\hat{V}_{i}$ does 
not anti-commute with $\Gamma^{\text{1D}}$ when the 
direction of the Zeeman field is along the $y$-direction. 
If we find an operator which anti-commutes with Hamiltonian, i.e. the system has a chiral symmetry,\cite{Schnyder08} we can define winding number by 
following procedures.\cite{wen,STYY11,Tewari2012L}
First, we diagonalize $\Gamma^{\text{1D}}$ by the unitary matrix $U_1$
\begin{equation}
U_1^{\dagger}\Gamma^{\text{1D}}U_1=\left(
    \begin{array}{cc}
      I_{2\times 2} &0 \\
     0& -I_{2\times 2} 
    \end{array}
  \right).
\end{equation}
Using this $U_1$, the Hamiltonian is transformed as 
\begin{equation}
U_{1}^{\dagger}\mathcal{H}_i(k_x)U_{1}=\left(
    \begin{array}{cc}
      0&q(k_x) \\
     q(k_x)^{\dagger} &0
    \end{array}
  \right),
\end{equation}
with $q(k_x)=(-2t\cos k_x-\mu)\sigma_0+2\lambda\sin k_x\sigma_y-i(\Delta_0+2\Delta_1\cos k_x)\sigma_y$. 
The winding number $w_{1d}$ is defined as 
\begin{equation}
w_{1d}=\frac{1}{2\pi i}\oint D^{-1}dD,\;\; D=\text{det}q(k_x), 
\end{equation}
where $\oint$ represents the line integral over closed loop in 1D Brillouin zone  
\cite{wen}. 
This quantity means how many times the trajectory of the 
det$q(k_x)=m_{1} + i m_{2}$ wraps around the origin in complex space, 
where $m_{1}$ and $m_{2}$ are real quantities.  
The number $w_{1d}$ is a topological invariant, which keeps its value unless the energy gap of the Hamiltonian is closed.
If $w_{1d}$ has a non-zero value, 
due to the bulk-edge correspondence, there are ZESs.\cite{wen}
As an example, we plot trajectories of det $q(k_x)$ in complex $m_{1}\text{-}m_{2}$ 
plane and the energy dispersion 
of $\mathcal{H}_d^{\text{1D}}(k_x)$ in the case of $s_{\pm}$-wave pairing 
without the Zeeman field (Figs. 13 (a) and (b))
and with it (Figs. 13 (c), (d) (e) and (f)). 
We can easily find $w_{1d}=2$ without the Zeeman field as shown in Fig. 13 (a). 
$w_{1d}$ remains to be 2 up to $V_{x}=V_{c1}$ ($V_{c1}/t\sim1.02$) 
which is the critical value of the bulk energy gap closure. 
At $V_{x}=V_{c1}$, the trajectory just crosses $(m_{1},m_{2})=(0,0)$ (Fig. 13(c))
and bulk energy gap closes at $k_{x}=0$ (Fig. 13(d)). 
If $V_{x}$ exceeds $V_{c1}$, the resulting $w_{1d}$ becomes 1. 
At $V_{x}=V_{c2}$, with $V_{c2}/t\sim3.06$, 
the trajectory crosses  $(m_{1},m_{2})=(0,0)$ (Fig. 13 (e)) and 
the bulk energy gap closes at $k_{x}=\pm \pi$ (Fig. 13(f)). 
Above $V_{c2}$, $w_{1d}$ becomes zero.  
These properties are consistent with that the topological invariant $w_{1d}$ 
changes when the bulk energy gap is closed.  
We also calculate $w_{1d}$ under the  Zeeman field in the $z$-direction  for $s_{\pm}$-wave pairing. 
The change of $w_{1d}$ and critical values of $V_{c1}$ and $V_{c2}$ are the 
same as in the case with the  Zeeman field in the $x$-direction. 
As a reference, we obtain $w_{1d}$ 
for  $s_{++}$-wave pairing  with the Zeeman field 
both in $x$ and  $z$-directions. 
In these directions, $w_{1d}$ changes 0, 1, 0 with the increase of 
the Zeeman field. The critical values 
$V_{c1}$ and $V_{c2}$ are given as 
$V_{c1} \sim 1.02$ and $V_{c2} \sim 3.00$, respectively. 
The above results are shown in Table I. 
The arrow in the Table 1 shows the change of the topological numbers 
with the increase of the Zeeman field. 
\begin{table}[htb]
  \begin{center}
    \begin{tabular}{|c|c|c|c|} \hline
      Pairing & $x$-direction &$y$-direction &$z$-direction \\ \hline \hline
      $s_{\pm}$ & $2\to1\to0$&$\times$ &$2\to1\to0$ \\
      $s_{++}$ & $0\to1\to0$&$\times$ &$0\to1\to0$ \\ \hline
    \end{tabular}
     \end{center}
    \caption{Changes of $w_{1d}$ in 1D with the increase of Zeeman field. The critical values of the Zeeman field where, the winding number is changed, are mentioned in Appendix. A.}
 \label{tab:price}
\end{table}

From this table, it is clear that the difference between $s_{\pm}$ and $s_{++}$-wave pairings is whether the state with $w_{1d}=2$ is realized or not
at the lower Zeeman field in the $x$ and $z$-direction.
This point is also clearly seen in Figs. 4 and  5.
Up to $V_x=V_{c1}$, ZES and resulting ZEP appear in the LDOS on the edge only for $s_{\pm}$-wave pairing.
The response of the $w_{1d}$ to the Zeeman field is seriously different between $s_{\pm}$ and $s_{++}$-wave pairing. \par

Next, we mention the anisotropic response of LDOS 
to the applied Zeeman field. 
To understand this, both the time-reversal and the mirror reflection symmetries are essential.
Suppose we have the Zeeman field term in the $x$-direction, 
$\hat{V}_{x}= V_x\sigma_x\tau_z$.
This term breaks both the time-reversal symmetry and the mirror reflection symmetry, 
\begin{equation}
\Theta^{\dagger}V_x\sigma_x\tau_z\Theta=-V_x\sigma_x\tau_z,
\end{equation} 
\begin{equation}
\mathcal{M}_{xz}^{\dagger}V_x\sigma_x\tau_z\mathcal{M}_{xz}=-V_x\sigma_x\tau_z.
\end{equation} 
However, there is the pseudo-time-reversal symmetry, i.e. the product of the time-reversal symmetry and the mirror reflection symmetry.
\begin{equation}
(\mathcal{M}_{xz}\Theta)^{\dagger}V_x\sigma_x\tau_z\mathcal{M}_{xz}\Theta=V_x\sigma_x\tau_z.
\end{equation}
The Zeeman field in the $z$-direction also has this combined symmetry.
\begin{table}[htb]
  \begin{center}
    \begin{tabular}{|c|c|c|c|} \hline
      Direction&TR&MR&TR+MR \\ \hline \hline
      $x$-direction & $\times$&$\times$ &$\bigcirc$ \\
      $y$-direction&$\times$ & $\bigcirc$ & $\times$ \\ 
z-direction & $\times$&$\times$ &$\bigcirc$  \\ \hline
    \end{tabular}
  \end{center}
\caption{Symmetries of the Hamiltonian with 
the Zeeman field in the $x$,$y$, and $z$-directions. 
TR, MR, and TR+MR mean 
the time-reversal, mirror reflection, and pseudo-time-reversal 
 symmetries, respectively.
The symbol $\times$ ($\bigcirc$)
represents that the corresponding symmetry is broken 
(not broken).  }
\label{tab:price}
\end{table}
In the presence of this symmetry, we can define topological number
and understand the origin of the edge state in the context of the bulk-edge correspondence. 
\par
Besides above discussions, 
we can provide another explanation about 
this anisotropic response to the Zeeman field by the index theorem.
We focus on the zero energy edge state and analyze them by the eigenstate of 
$\Gamma^{\text{1D}}$. 
It is known that the index theorem guarantees the following relations 
\cite{TMYYS10} as 
\begin{equation}
w_{1d}=n_0^{+}-n_0^{-}\label{plus}, 
\end{equation}
or
\begin{equation}
w_{1d}=n_0^{-}-n_0^{+}\label{minus},
\end{equation} 
where, $n_0^{+(-)}$ is a number of eigenstate of $\Gamma^{\text{1D}}$ whose eigenvalue is $+(-)1$.  

Suppose Eq. (\ref{plus}) holds and $w_{1d}=2$. ZES can be written as $(\xi^{(a)}_{\uparrow},\xi^{(a)}_{\downarrow},\xi^{*(a)}_{\uparrow},\xi^{*(a)}_{\downarrow})^{T}\ (a=1,2)$ by using particle-hole symmetry. From $n_0^{+}=2$, it gives
\begin{equation}
\Gamma^{\text{1D}}
\left(
\begin{array}{c}
      \xi^{(a)}_{\uparrow}  \\
\xi^{(a)}_{\downarrow}\\
     \xi^{*(a)}_{\uparrow}\\
\xi^{*(a)}_{\downarrow}\\
    \end{array}
  \right)=(+1)\left(
\begin{array}{c}
      \xi^{(a)}_{\uparrow}  \\
\xi^{(a)}_{\downarrow}\\
     \xi^{*(a)}_{\uparrow}\\
\xi^{*(a)}_{\downarrow}\\
    \end{array}
  \right)\ (a=1,2).
\end{equation} 
Therefore,  following relations are satisfied for the ZES
\begin{equation}
 \xi^{(a)}_{\uparrow}= \xi^{*(a)}_{\uparrow}\ ,\  \xi^{(a)}_{\downarrow}= \xi^{*(a)}_{\downarrow}\label{kankei}. 
\end{equation}

If we expand the field operator  of 1D Rashba semiconductor nanowire $\mathbf{\Psi}=(\hat{\psi}_{\uparrow},\hat{\psi}_{\downarrow},\hat{\psi}^{\dagger}_{\uparrow},\hat{\psi}^{\dagger}_{\downarrow})^{T}$ 
only by remaining ZES ignoring the high energy state, 
we obtain
\begin{equation}
\mathbf{\Psi}=\sum_{a=1,2}\gamma^{(a)}\left(
\begin{array}{c}
      \xi^{(a)}_{\uparrow}  \\
\xi^{(a)}_{\downarrow}\\
     \xi^{*(a)}_{\uparrow}\\
\xi^{*(a)}_{\downarrow}\\
    \end{array}
  \right)\label{kankei2}.
\end{equation}
with real Majorana operator $\gamma^{(a)}$ satisfying 
$\gamma^{(a)\dagger}=\gamma^{(a)}$.  
We can introduce spin operator as follows:\cite{spin}
\begin{equation}
S_{\alpha}\equiv \frac{1}{4}\bigl[\hat{\psi}_i(\sigma_{\alpha})_{ij}\hat{\psi}^{\dagger}_j-\hat{\psi}^{\dagger}_i(\sigma^{T}_{\alpha})_{ij}\hat{\psi}_j\bigr]
\end{equation}
From Eqs. (\ref{kankei}) and (\ref{kankei2}), we can  find that $S_x=0$, 
$S_z$=0, and $S_y$$\neq0$. 
This means that the  spin of zero energy Majorana edge state 
is not coupled with the Zeeman field applied in the $x$, $z$-directions but 
can be coupled with that in the $y$-direction.
It is an Ising like character of Majorana edge state 
\cite{spin,Shindo,Nakosai2}. 
This result is consistent  with the anisotropic 
response of the Zeeman field discussed in previous section. 
\subsection{2D case}\label{sec4b}
In 2D case, the chiral operator defined in 1D, $\Gamma^{\text{1D}}$ 
satisfies
\begin{equation}
(\Gamma^{\text{1D}})^{\dagger}\mathcal{H}_{i}^{2D}(k_x,k_y)\Gamma^{\text{1D}}=-\mathcal{H}_{i}^{2D}(k_x,-k_y).
\end{equation}
Therefore, only at $k_y=0,\pi$, $\mathcal{H}_{i}^{2D}(k_x,k_y)$ anti-commutes with $\Gamma^{\text{1D}}$. 
We can introduce a winding number by similar way in 1D for 
fixed $k_{y}=0$ or $k_{y}=\pi$. 
The resulting winding number at $k_y=0(k_y=\pi)$ reads
\begin{equation}
w_{1d}(0(\pi))=
\frac{1}{2\pi i}\oint D^{-1}dD,\;\; D=\text{det}q(k_x,0(\pi)),
\end{equation} 
where, $q(k_x,k_y)=(-2t\cos k_x-2t\cos k_y-\mu)\sigma_0+2\lambda\sin k_x\sigma_y-i(\Delta_0+2\Delta_1\cos k_x+2\Delta_1\cos k_y)\sigma_y$. 
The calculated winding numbers at $k_y=0,\pi$ both in the case of $s_{\pm}$ and $s_{++}$-wave pairing  is shown in Table III and IV with  
the applied Zeeman in the $x$ and $z$-directions, respectively.   

\begin{table}[htb]
  \begin{center}
    \begin{tabular}{|c|c|} \hline
      Pairing & $(w_{1d}(0),w_{1d}(\pi))$ \\ \hline \hline
      $s_{\pm}$ &$(2,0)\to(1,1)\to(0,1)\to(0,0)$\\

      $s_{++}$ & $(0,0)\to(1,1)\to(0,1)\to(0,0)$ \\ \hline
    \end{tabular}
  \end{center}
    \caption{Changes of $(w_{1d}(0),w_{1d}(\pi))$ with the Zeeman field in $x$-direction in 2D. The critical values of the Zeeman field where winding number is changed are mentioned in Appendix.A}
    \label{2Dx}
\end{table}

\begin{table}[htb]
  \begin{center}
    \begin{tabular}{|c|c|} \hline
      Pairing & $(w_{1d}(0),w_{1d}(\pi))$ \\ \hline \hline
      $s_{\pm}$ &$(2,0)\to(1,1)\to(0,1)\to(0,0)$\\

      $s_{++}$ & $(0,0)\to(1,1)\to(0,1)\to(0,0)$ \\ \hline
    \end{tabular}
  \end{center}
    \caption{Changes of $(w_{1d}(0),w_{1d}(\pi))$ with the Zeeman field in $z$-direction in 2D. The critical values of the Zeeman field where winding number is changed are mentioned in Appendix.A}
    \label{2Dz}
\end{table}
The arrows in the tables show the change of the topological numbers 
with the increase of the Zeeman field. 
We have confirmed that when the winding number is changed, 
the energy gap of the Hamiltonian at $k_y=0$ or $k_y=\pi$ is closed. 
For $s_{\pm}$-wave pairing, there are three 
 critical field $V_{c1}$, $V_{c2}$, and $V_{c3}$, 
where the energy gap at $k_y=0$ or $k_y=\pi$ closes between 
$(w_{1d}(0),w_{1d}(\pi))=(2,0)$ and (1,1), 
(1,1) and (0,1), and 
(0,1) and (0,0), respectively. 
These are $V_{c1} \sim 1.02$, $V_{c2} \sim 3.06$ and $V_{c3} \sim 5.10$
both for $x$-direction and $z$-direction cases. 
For $s_{++}$-wave pairing, 
the corresponding $V_{c1}$, $V_{c2}$, and $V_{c3}$ 
given by $V_{c1} \sim 1.02$, $V_{c2} \sim 3.00$ and $V_{c3} \sim 5.00$
both for $x$-direction and $z$-direction cases. 
The $w_{1d}$ changes (0,0), (1,1), (0,1) and (0,0) with the increase of Zeeman field. 
We can see that the results shown in Table III are consistent 
with the change of 
the number of the chiral edge modes at $k_y=0$ and $k_{y}=\pi$ 
shown in Figs. 7 and 10. 
The difference between $s_{\pm}$ and $s_{++}$-wave 
only appears at the lower Zeeman field 
whether the topological state with 
$(w_{1d}(0),w_{1d}(\pi))=(2,0)$ is realized or not. 
In the higher Zeeman field, 
since the number of the  Fermi surface becomes one   
and the sign change of the pair potential does not occur any more.   
Thus, there is no essential difference between 
$s_{\pm}$ and $s_{++}$-wave pairings. 
\section{Conclusion}\label{sec5}
\label{summary}
In this paper, 
we have studied the Majorana edge states 
and their topological properties 
of one-dimensional(1D) and two-dimensional(2D) Rashba semiconductor 
deposited on iron-based superconductor 
under the applied Zeeman field for various directions. 
Using the recursive Green's
function method, we have calculated the LDOS and ARLDOS  both for 
$s_{\pm}$ and $s_{++}$-wave pairings. 
We have discussed whether Majorana edge state emerges or not based on the 
symmetry of the Hamiltonian. 
We have clarified Majorana edge states are protected by the mirror reflection symmetry and therefore it shows 
an anisotropic response to the directions of the applied Zeeman field.  
The resulting LDOS and winding numbers  for $s_{\pm}$ and $s_{++}$-wave pairings are essentially different at the lower Zeeman field. 
These results serve as a guide to determine the pairing symmetry of 
iron-pnictide.

\section{Acknowledgments}
We would like to thank M. Sato for valuable discussions. 
This work was supported in part by a Grant-in Aid
for Scientific Research from MEXT of Japan, "Topological
Quantum Phenomena", Grants No. 22103005 and  
EU-Japan program "IRON SEA".

\appendix
\section{Critical values of the Zeeman field}
First, we focus on the 1D case. 
The eigenvalue energy of the Hamiltonian with the Zeeman field in the $i$-direction ($i$=$x,z$) is given by
\begin{eqnarray}
E(k_x)=\pm\sqrt{\alpha(k_x)\pm2\sqrt{\beta(k_x)}}\notag\\
\alpha(k_x)=\xi_{k_x}^2+|\Lambda(k_x)|^2+V_{i}^2+|\Delta(k_x)|^2\notag\\
\beta(k_x)=\xi_{k_x}^2|\Lambda(k_x)|^2+(\xi_{k_x}^2+|\Delta(k_x)|^2)V_{i}^2,
\end{eqnarray}
where, $\xi_{k_x}=-2t\cos k_x-\mu$, $\Lambda(k_x)=-i2\lambda_R\sin k_x$, $\Delta(k_x)=\Delta_0+2\Delta_1\cos k_x$. The conditions of the gap closure is given as\cite{sato06}
\begin{equation}
\label{acon}
\xi_{k_x}^2+|\Delta(k_x)|^2=V_{i}^2+|\Lambda(k_x)|^2,\;\;|\Delta(k_x)||\Lambda(k_x)|=0
\end{equation}
From the second equation of \eqref{acon}, we obtain $k_x=0$, $\pi$, $\text{Cos}^{-1}(-\Delta_0/2\Delta_1)$ and put this $k_x$ into the first equation, 
then we get the value of $V_{i}$. 
In the actual numerical calculations, 
we choose $t$=1, $\lambda_R=0.5$, $\mu=-1$.
$\Delta_0=-0.2$ and $\Delta_1=0.2$ for $s_{\pm}$-wave pairing 
$\Delta_0=0.2$ and $\Delta_1=0.0$ for $s_{++}$-wave pairing. 
\par
We obtain the critical values of the Zeeman field for the gap closure. As for the $s_{\pm}$-wave case, critical values $V_i=V_{c1}$, $V_i=V_{c2}$ are $V_{c1}/t\sim1.02$, $V_{c2}/t\sim3.06$. As for the $s_{++}$-wave case, corresponding critical values are $V_{c1}/t\sim1.02 $ and $V_{c2}/t\sim3.00 $. 
Also, we confirm that at these critical values, winding number has a jump.

\par In 2D case, the eigenvalue energy of the Hamiltonian at $k_y=0(\pi)$ with the Zeeman field 
in the $i$-directions ($i$=$x,z$)reads

\begin{eqnarray}
&E(k_x,0(\pi))=\pm\sqrt{\alpha(k_x,0(\pi))\pm2\sqrt{\beta(k_x,0(\pi))}}\notag\\
&\alpha(k_x,0(\pi))=\xi_{k_x,0(\pi)}^2+|\Lambda(k_x,0(\pi))|^2+V_{i}^2+|\Delta(k_x,0(\pi))|^2\notag\\
&\beta(k_x,0(\pi))=\xi_{k_x,0(\pi)}^2|\Lambda(k_x,0(\pi))|^2\notag\\
&+(\xi_{k_x,0(\pi)}^2+|\Delta(k_x,0(\pi))|^2)V_{i}^2,
\end{eqnarray}
where,
\begin{eqnarray}
\xi_{k_x,0(\pi)}=-2t\cos k_x-2t\cos(0(\pi))-\mu, \\
\Lambda(k_x,0(\pi))=-i2\lambda_R\sin k_x+2\lambda_R\sin(0(\pi)), \\
\Delta(k_x,0(\pi))=\Delta_0+2\Delta_1\cos k_x+2\Delta_1\cos (0(\pi)).
\end{eqnarray}
 The condition of gap closure is as follows.
\begin{eqnarray}
\label{2Dcon}
\xi_{k_x,0(\pi)}^2+|\Delta(k_x,0(\pi))|^2=V_{z}^2+|\Lambda(k_x,0(\pi))|^2\notag\\
|\Delta(k_x,0(\pi))||\Lambda(k_x,0(\pi))|=0
\end{eqnarray}
For the same set of parameters as in 1D case, we get the critical values for the gap closure at $k_y=0,\pi$. In the case of $s_{\pm}$, critical values $V_i=V_{c1}$, $V_i=V_{c2}$ and $V_i=V_{c3}$ are given by
$V_{c1}/t\sim1.02$, $V_{c2}/t\sim3.06$, and $V_{c3}/t\sim5.10$, while in the case of $s_{++}$, $V_{c1}/t\sim1.02$, $V_{c2}/t\sim3.00$, and $V_{c3}/t\sim5.00$.   
We also confirm that at these values, the winding number is changed.

\end{document}